\documentclass{Interspeech}
\usepackage{colortbl}
\usepackage{hyperref}
\usepackage[table]{xcolor}

\interspeechcameraready

\title{Adaptive Differential Denoising for Respiratory Sounds Classification}

\author[affiliation={1}]{Gaoyang}{Dong}
\author[affiliation={2}]{Zhicheng}{Zhang}
\author[affiliation={3*}]{Ping}{Sun}
\author[affiliation={1*}]{Minghui}{Zhang}

\affiliation{School of Information Engineering}{Nanchang University}{China}
\affiliation{School of Computer Science and Technology}{Fudan University}{China}
\affiliation{}{1st Affiliated Hospital of Nanchang University}{China}
\email{donggyncu@email.ncu.edu.cn, zhangminghui@ncu.edu.cn}


\usepackage{comment}
\newcommand\blfootnote[1]{%
  \begingroup
  \renewcommand\thefootnote{}\footnote{#1}%
  \addtocounter{footnote}{-1}%
  \endgroup
}

\begin{document}

\maketitle

\begin{abstract}
    
   Automated respiratory sound classification faces practical challenges from background noise and insufficient denoising in existing systems. 
    We propose Adaptive Differential Denoising network, that integrates noise suppression and pathological feature preservation via three innovations: 
    1) Adaptive Frequency Filter with learnable spectral masks and soft shrink to eliminate noise while retaining diagnostic high-frequency components; 
    2) A Differential Denoise Layer using differential attention to reduce noise-induced variations through augmented sample comparisons; 
    3) A bias denoising loss jointly optimizing classification and robustness without clean labels. 
    Experiments on the ICBHI2017 dataset show that our method achieves 65.53\% of the Score, which is improved by 1.99\% over the previous sota method.
    The code is available in https://github.com/deegy666/ADD-RSC
\end{abstract}

\section{Introduction}

Automated respiratory sound classification (ARSC) has emerged as a critical tool for the non-invasive diagnosis of pulmonary diseases such as asthma, pneumonia, and chronic obstructive pulmonary disease (COPD) \cite{pramono2017automatic}. By analyzing auscultation signals captured via digital stethoscopes or wearable sensors, ARSC systems aim to categorize pathological patterns (e.g., crackles, wheezes) with clinical-level accuracy. This capability is particularly valuable for telemedicine applications and early screening in resource-constrained regions with limited access to specialized pulmonologists \cite{haleem2021telemedicine}\blfootnote{*Corresponding author. And this research is supported by the National Natural Science
Foundation of China under Grant No. 82260024}.

Despite its clinical promise, current ARSC algorithms encounter two major challenges associated with data limitations: the scarcity of annotated samples and commonly existing environmental noise.
Recently, researchers have sought to mitigate data scarcity through novel augmentation techniques. Siddhartha Gairola et al. proposed blank region clipping and concatenation-based augmentation to preserve pathological semantics while expanding spectral diversity \cite{gairola2021respirenet}. 
June-Woo Kim et al. developed RepAug to extract augmented features from pre-trained models on speech waveforms \cite{kim2024repaugment}. The use of adversarial fine-tuning techniques was also introduced by June-Woo Kim et al. as a data augmentation approach  \cite{Ad-F-tuning}. 
Concurrently, deep learning architectures like Convolutional Neural Network(CNN) \cite{gairola2021respirenet,chang2022example,wang2022domain}, Audio Spectrogram Transformer(AST) \cite{kim2024repaugment,sg-scl,patch-mix,dong2025respiratory} and Hierarchical Token-Semantic Audio Transformer(HTS-AT) \cite{swim, wang2024lightweight} have shown remarkable efficacy in feature extraction by capturing long-range dependencies in respiratory acoustics.

However, the issue of noise contamination remains a major challenge in respiratory sound classification (RSC). Clinical recordings are often affected by non-stationary disturbances arising from limb movements (e.g., friction artifacts, cardiac sounds), ambient noise (e.g., conversational interference, stethoscope-clothing contact noise), and transducer-related artifacts.
Conventional denoise methods such as spectral subtraction \cite{upadhyay2015speech} and the Wiener filter technique \cite{ramasubramanian2008speech} often fail to effectively suppress these disturbances without compromising diagnostically critical high-frequency components. Commonly used techniques, such as Butterworth bandpass filtering \cite{podder2020design} and Wavelet decomposition \cite{ali2015improved}, impose rigid spectral constraints: the former excessively attenuates frequencies beyond the standard respiratory range (100-2000 Hz), inadvertently discarding pathological features such as high-frequency crackles ($>$2000 Hz) \cite{miller2022physiological}, whereas the latter leads to transient feature loss due to its reliance on fixed basis functions.

While deep learning has driven advances in respiratory sound analysis, 
most existing methods follow a sequential denoising-classification paradigm \cite{tzeng2024improving,zhao2023mossformer,luo2019conv}, which presents three fundamental limitations.
Firstly, the absence of dedicated respiratory sound denoising datasets compels researchers to utilize generic audio denoisers trained on non-biomedical recordings. It creates domain mismatches that compromise the preservation of pathological features.
Secondly, explicit denoising modules are optimised for general audio tasks. Yet, these modules often unintentionally eliminate subtle pathological features essential for fine-grained classification.
Thirdly, the cascaded structure fundamentally decouples the objectives of denoising and classification, thereby hindering the joint optimization of noise-invariant representations.
This approach is not effective when dealing with non-stationary physiological noise. Such noise, for example, heartbeat interference, shares similar spectral-temporal features with the target respiratory event, making its removal particularly challenging.

To bridge this gap, this paper introduces an implicit respiratory sound denoising approach that eliminates the need for the traditional sequential denoising-classification paradigm, enabling direct application to RSC. Overall, our contribution is threefold:
\begin{itemize}
\item To the best of our knowledge, this study is the first to explore deep learning-based denoising techniques specifically for respiratory sounds.

\item We propose an Adaptive Differential Denoising (ADD) network with an adaptive frequency filter to suppress exogenous interference and a differential attention module to refine subtle breathing noise. Furthermore, the model is optimized using a bias-aware loss function, which enhances classification performance through implicit denoising.

\item Our model achieves state-of-the-art performance on the ICBHI 2017 benchmark, surpassing previous best methods by 1.99\%.

\end{itemize}

\begin{figure*}[t]
  \centering
  \includegraphics[width=0.9\linewidth]{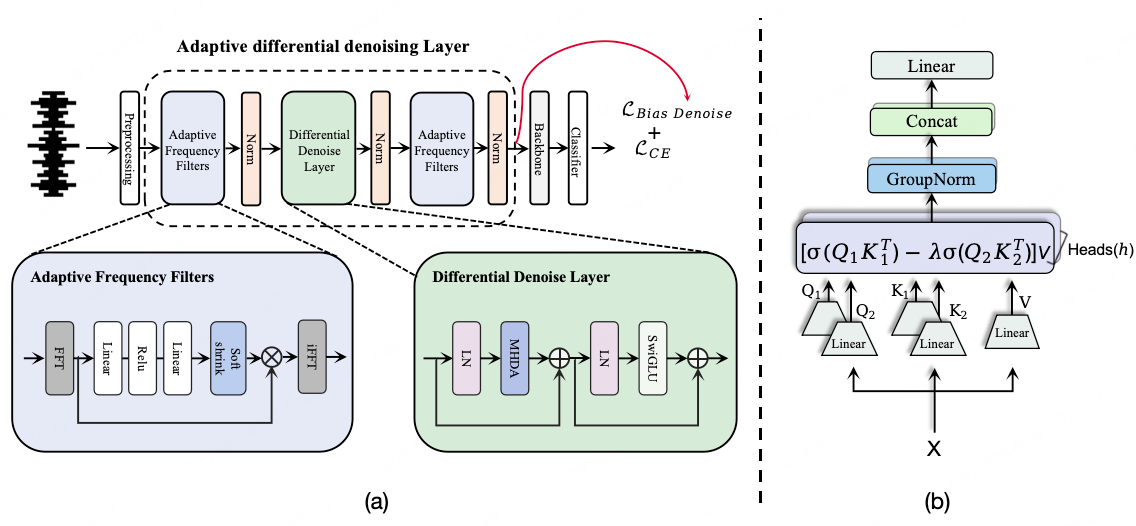}
  \caption{Overall illustration of proposed model architecture. (a) shows Adaptive Differential Denoising Framework and (b) shows the specific
details of the Multi-Head Differential Attention
(MHDA) Mechanism.}
  \label{fig:1}
\end{figure*}

\section{Proposed Methods}

In this section, we introduce the Adaptive Differential Denoising framework for RSC. The overall pipeline is illustrated in Figure \ref{fig:1}, which consists of three key components: (1) Adaptive Frequency Filters (AFF), (2) Differential Denoise Layer (DDL), and (3) Loss Functions. Below, we describe each component in detail.

\subsection{Preprocessing}

Before feeding the audio signals into the model, we performed a series of preprocessing steps to standardize the input format and enhance model robustness. Given that the respiratory sound recordings in the original datasets were collected using different devices with varying sampling rates ranging from 4 kHz to 44.1 kHz, all audio signals were resampled to 16 kHz.

In the ICBHI dataset, the duration of respiratory cycles varies significantly, ranging from 0.2 seconds to 16.2 seconds. This variability poses challenges for training neural networks, which require fixed-size inputs. We determined that an 8-second length yields optimal performance. To ensure consistency, shorter recordings were extended using cyclic padding, while longer recordings were truncated accordingly. Additionally, amplitude normalization was applied to maintain uniform input scaling. Finally, we transformed the audio signals into the Mel spectrogram($\mathbf{X} (t, f)$), which was computed with 64 Mel filters and a window size of 1024 samples.

\subsection{Adaptive Frequency Filters}

We use the Fast Fourier Transform (FFT) to convert $\mathbf{X} (t, f)$ from the time-frequency domain to the spectral domain, so as to extract the corresponding frequency representation. The detailed transformation formula is as follows:
\begin{equation}
    \mathbf{X} (U, V) = \sum_{t=0}^{T-1} \sum_{f=0}^{F-1} \mathbf{X} (t, f) e^{-2\pi i(Ut+Vf)}.
\end{equation}
where 
$\mathbf{X} (U, V)$ represents the Modulation Spectrum, $U$ and $V$ correspond to the modulation frequencies along the temporal and spectral dimensions, respectively.
$F$ represents the size of the frequency dimension of the spectrogram and $T$ represents the size of the time dimension of the spectrogram. 

Subsequently, according to the Convolution Theorem, convolution in the time domain corresponds to element-wise multiplication in the frequency domain.
We apply a filtering mechanism with learnable instance adaptive masks \cite{huang2023adaptive}. The mechanism is used for the frequency representation of the signal and is designed for efficient and adaptive denoising of breath sounds.
It is expressed as follows:
\begin{equation}
    {\hat{X}(t, f)} = \mathcal{F} ^{-1} [S_\alpha(\mathcal{M}(\mathbf{X} (U, V))) \odot \mathbf{X} (U, V)],
\end{equation}
where $\mathcal{M}(\cdot)$ learnable instance adaptive masks.
Specifically, it consists of learnable Linear layer, Relu, and another Linear layer. $\mathcal{F}^{-1}(\cdot)$ denotes inverse Fourier transform. $S_\alpha(\cdot)$ is soft shrink, and is represented as follows:
\begin{equation}
S_\alpha(x)=\mathrm{sign}(x)\max\{|x|-\alpha,0\}
\end{equation}
where $\alpha$ is a tuning parameter that controls the sparsity. In addition to noise reduction, it is also possible to improve sparsity and deal with L1 normalisation.

\subsection{Differential Denoise Layer}

To eliminate subtle noise while preserving key respiratory sound features, we introduce the Differential Denoise Layer. Unlike conventional denoising techniques which depend on explicit noise estimation, our method leverages contrastive feature refinement by computing differential attention between multiple augmented views of the input. This approach effectively suppresses noise while retaining discriminative signal components.

As illustrated in Figure 1(a), the DDL initially applies Layer Normalization (LN) to stabilize feature distributions and mitigate variations caused by diverse recording conditions. The core of the DDL relies on the Multi-Head Differential Attention (MHDA) mechanism, which captures feature variations between two perturbed representations of the input signal \cite{ye2024differential}. The MHDA details are shown in Figure 1(b). Specifically, given two feature sets , MHDA computes a contrastive attention score as follows:
\begin{equation}
    [\mathbf{Q}_1; \mathbf{Q}_2] = \mathbf{XW}^Q, \hspace{0.3em} [\mathbf{K}_1; \mathbf{K}_2] = \mathbf{XW}^K,\hspace{0.3em} \mathbf{V} = \mathbf{XW}^V
\end{equation}

\vspace{-0.5em}

\begin{equation}
    \mathbf{A} = \left(\sigma\left(\frac{\mathbf{Q}_1 \mathbf{K}_1^T}{\sqrt{d}}\right) - \lambda \cdot \sigma\left(\frac{\mathbf{Q}_2 \mathbf{K}_2^T}{\sqrt{d}}\right)\right) \mathbf{V} \label{eq:diffatten}
\end{equation}

where \( \mathbf{W}^Q, \mathbf{W}^K, \mathbf{W}^V \) are learnable projection matrices. The queries and keys are divided into \( \mathbf{Q}_1, \mathbf{Q}_2 \) and \( \mathbf{K}_1, \mathbf{K}_2 \), allowing for the computation of two separate attention maps. As shown in Equation \ref{eq:diffatten}, \(\sigma\) denotes the softmax function, and \(\lambda\) is a learnable scaling factor that controls the contribution of the second attention term. By subtracting two attention maps, MHDA filters out noise-sensitive variations while preserving stable, high-confidence features critical for classification.

Following the attention computation, a subsequent Layer Normalization stage further refines the denoised representation. Finally, SwishGLU activation introduces non-linearity while preserving smooth gradient flow, thereby ensuring the retention of discriminative features.

By leveraging differential attention mechanisms, the DDL suppresses noise-sensitive perturbations and enhances the model’s ability to focus on stable respiratory sound patterns. This approach significantly improves robustness against background noise and domain shifts, making it particularly effective for real-world respiratory sound classification.

\subsection{Loss Functions}

The proposed framework utilizes a hybrid loss function composed of two principal components:

\begin{equation} \label{eq:total_loss}
\mathcal{L} = \beta \mathcal{L}_{\text{Bias Denoise}} + (1-\beta) \mathcal{L}_{\text{CE}}
\end{equation}

where $\beta$ is the hyperparameter, which controls the trade-off between denoising guidance and classification accuracy.

We use label smoothing cross-entropy to regularize the denoising process, as expressed by the following equation:
\begin{equation}
    {\mathcal{L}_\text{Bias Denoise}} =  - \sum\limits_i \left[y_c (1 - \epsilon)  + \frac{\epsilon}{C}\right] \log \left[\phi(Norm(p)\right] 
\end{equation}
where $\phi$ is the network for mapping the previously denoised features, specifically we use a 1 $\times$ 1 convolutional layer to represent it, $y_c$ is the true label, $C$ is the total number of classifications, and $p$ is the feature derived after the DDL network.

Unlike spectral distance measures that require clean reference signals, our formulation only relies on categorical labels-crucial for real-world deployment where paired clean/noisy data is unavailable. The smoothing operation injects prior medical knowledge about pathological feature distributions into the denoising process. Furthermore, clinical noise frequently overlaps spectrally with target signals (e.g., heart sounds masking fine crackles). The label smoothing parameter $\epsilon$ acts as an uncertainty buffer, preventing overconfident removal of ambiguous frequency bands that may contain diagnostic information.

\section{Experiments}

\subsection{Experimental Setup}

\subsubsection{Dataset}

This experiment employs the ICBHI 2017 dataset \cite{icbhi}, the largest publicly available respiratory sound corpus, for evaluation. Audio data were collected from 126 participants. Due to the complexity of the clinical recording environment, the original recordings may contain a variety of real noise disturbances, such as ambient background sounds (e.g., ward equipment humming), physiological disturbances (e.g., heart sound aliasing), and contact noise from stethoscope rubbing against clothing.
The overlapping nature of these noises with the respiratory sound in the time-frequency domain significantly increases the challenge of the classification task.
The dataset was recorded using a variety of devices with different sampling rates (4 kHz, 10 kHz, 44.1 kHz), and the duration of a single recording ranged from 10 to 90 seconds. The heterogeneity of the noise makes this dataset effective for validating the robustness of the model in real scenarios. According to the official protocol, the dataset is partitioned into training and test sets (60\%-40\%) while ensuring no patient identity overlap to prevent data leakage.

\subsubsection{Evaluation Metrics}

To ensure the accuracy and consistency of our assessments, we have strictly followed the official evaluation methodology set out by the ICBHI 2017 Challenge. The methodology is comprehensive and detailed, which consists of three key metrics: Sensitivity ($Se$), Specificity ($Sp$) and the $Score$. The $Se$, $Sp$, and $Score$ are defined as follows:

\begin{equation}
Se=\frac{\#of\ Correctly\ Recognised\ Abnormal\ Sounds}{\#of\ Total\ Abnormal\ Sounds}
\end{equation}

\begin{equation}
Sp=\frac{\#of\ Correctly\ Recognized\ Normal\ Sounds}{\#of\ Total\ Normal\ Sounds}
\end{equation}

\begin{equation}
Score=\frac{Se+Sp}2
\end{equation}

where, $Se$ denotes the accuracy of recognising abnormal sounds among all abnormal sounds, while $Sp$ denotes the accuracy of recognising normal sounds among all actual normal sounds. $Score$ is averaged over the abnormal and normal sounds.
In particular, abnormal sounds contained Wheeze, Crackle and both (Wheeze\&Crackle) in this dataset 

\subsubsection{Training Details}
In the training phase, we conducted experiments using the Pytorch 2.3.1 framework on a Tesla V100 GPU. We used the Adam optimiser with weight decay set to 0.1, learning rate set to 5e-5, batch size 8, and epoch set to 50. For the Backbone network, we used the Resnet50 and AST-Base model pre-trained on the AudioSet dataset.
The experimental hyperparameters are specified: the $\beta$ is 0.5, the $\alpha$ is 0.02, the $\epsilon$ is 0.2.

\subsection{Experimental Results}
\subsubsection{Comparison with the sota methods}

\begin{table}[ht]
\centering
\caption{Comparison of Ours with state-of-the-art methods on ICBHI dataset. We compared it to previous works (the results from \cite{paperswithcode_icbhi_sota})}
\label{tab:result1}
\resizebox{\linewidth}{!}{
\begin{tabular}{l|ccc}
\toprule
\toprule
Model & \textit{$S_p$}(\%) & \textit{$S_e$}(\%) & \textit{Score}(\%)  \\

\midrule
LungRN+NL\cite{ma2019lungbrn} & 63.20 & 41.32 & 52.26 \\
RespireNet \cite{gairola2021respirenet} & 72.30 & 40.10 & 56.20 \\ 
Wang \textit{et al.} \cite{wang2022domain} (Splice)  & 70.40 & 40.20 & 55.30 \\
StochNorm\cite{nguyen2022lung}& 78.86 & 36.40 & 57.63 \\
CoTuning \cite{nguyen2022lung}& 79.34 & 37.24 & 58.29 \\
Chang \textit{et al.} \cite{chang22h_interspeech} & 69.92 & 35.85 & 52.89 \\
SCL \cite{moummad2022supervised}& 75.95 & 39.15  & 57.55 \\
\midrule
\rowcolor[gray]{0.75}\textbf{Ours}(Resnet50) & 83.76 & 34.18 & 58.97\\

\midrule	
AFT on Mixed-500 \cite{Ad-F-tuning} & 80.72 & 42.86 & 61.79 \\
AST Fine-tuning \cite{patch-mix} & 77.14 & 41.97 & 59.55 \\
Patch-Mix CL\cite{patch-mix} & 81.66 & 43.07 & 62.37 \\
M2D  \cite{m2d}& 81.51 &  45.08 & 63.29 \\
DAT \cite{sg-scl}& 	77.11 & 42.50 &  59.81 \\
SG-SCL \cite{sg-scl}  & 79.87 &43.55 &61.71		 \\
RepAugment \cite{kim2024repaugment} & 82.47 & 40.55 & 61.51 \\
BTS \cite{bts}& 81.40 & 45.67 & 63.54 \\
MVST \cite{he2024multi} &80.6 &44.39 &62.50 \\
LungAdapter \cite{xiao2024lungadapter} & 80.43 & 44.37 & 62.40 \\
CycleGuardian \cite{CycleGuardian} & 82.06 & 44.47 & 63.26 \\
\midrule
\rowcolor[gray]{0.75}\textbf{Ours}(AST) & \textbf{85.13} & \textbf{45.94} & \textbf{65.53}\\
\bottomrule
\bottomrule
\end{tabular}}
\end{table}

We compared our approach with various SOTA methods on the ICBHI 2017 dataset, as shown in Table \ref{tab:result1}. The table is divided into two parts: the upper section presents methods based on CNN, while the lower section focuses on Transformer-based models. 

In the upper part, for CNN-based approaches, our proposed approach to backbone networks using Resnet50 achieves a leading \textit{Score} of 58.97\%, surpassing all previous CNN variants (e.g., 2.68\% higher than CoTuning) through effective integration of adaptive differential denoising. This verifies our framework's compatibility with convolutional operations.

In the lower part of the table, when implemented on the AST Transformer backbone. Our method attains SOTA performance with 65.53\% $Score$ - outperforming all existing Transformer-based methods. 
Specifically, it exceeds the previous best Transformer approach (BTS [63.54\%]) by 1.99\%  improvement in $Score$ while simultaneously achieving the highest sensitivity ($S_e$=45.94\%, +0.27\% over BTS).
In particular, our $S_p$ reaches 85.13\%, establishing a 2.66\% margin over RepAugment's previous record (82.47\%), which demonstrates enhanced specificity without compromising detection sensitivity.

The consistent superiority across architectural paradigms (CNN/Transformer) strongly validates the architecture-agnostic nature of our adaptive differential denoising mechanism. In particular, simultaneous improvements in both metrics (traditionally exhibiting trade-off characteristics) suggest that our method effectively disentangles signal patterns from noise interference in respiratory sound analysis.

\subsubsection{Ablation Study}

\begin{table}[ht]
\centering
\caption{Ablation study of our proposed method on ICBHI 2017 dataset. The backbone is the AST.}
\label{tab:abl}

\resizebox{\linewidth}{!}{
\begin{tabular}{l|ccc}
\toprule
\toprule
Model & \textit{$S_p$}(\%) & \textit{$S_e$}(\%) & \textit{Score}(\%)  \\

\midrule
w AFF & 82.47 & 44.47 & 63.47 \\
w DDL & 83.73 & 44.03 & 63.88 \\ 
\midrule
w AFF+DDL & 84.41 & 44.78 & 64.60\\
w AFF+ $\mathcal{L}_{Bias Denoise}$ & 83.76 & 44.62 & 64.19 \\
w DDL+ $\mathcal{L}_{Bias Denoise}$ & 84.54 & 44.32 & 64.43\\
\midrule
\textbf{Ours} & \textbf{85.13} & \textbf{45.94} & \textbf{65.53}\\
\bottomrule
\bottomrule
\end{tabular}}
\end{table}

To evaluate the contributions of different components in our proposed method, we conduct an ablation study on the ICBHI 2017 dataset using AST as the backbone. The results are presented in Table \ref{tab:abl}.

We first examine the impact of the AFF module and DDL separately. Adding AFF improves $S_p$ to 82.47\% and achieves a \textit{Score} of 63.47\%, while incorporating DDL leads to an improved $S_p$ of 83.73\% and a \textit{Score} of 63.88\%. This demonstrates that both components enhance model performance, with DDL playing a more significant role in noise suppression and feature refinement.

Next, we analyze the performance of the model when AFF and DDL are combined. The model with both AFF and DDL achieves $S_p$ = 84.41\%, $S_e$ = 44.78\%, and a \textit{Score} of 64.60\%, showing that the two modules are complementary in improving classification performance. Furthermore, we explore the impact of integrating the bias denoise loss $\mathcal{L}_{Bias Denoise}$. The models using AFF + $\mathcal{L}_{Bias Denoise}$ and DDL + $\mathcal{L}_{Bias Denoise}$ achieve \textit{Score} of 64.19\% and 64.43\%, respectively, indicating that bias-aware denoising further refines feature representations and improves model robustness.

Finally, our full model, which integrates AFF, DDL, and $\mathcal{L}_{Bias Denoise}$, achieves the best performance, with $S_p$ = \textbf{85.13\%}, $S_e$ = \textbf{45.94\%}, and a final \textit{Score} of \textbf{65.53\%}. These results confirm that each component in our method plays a crucial role in enhancing performance, and their integration yields optimal classification accuracy on the ICBHI dataset.

\section{Conclusion}

In this work, we alleviate the existing gap in respiratory sound denoising. We introduce a novel end-to-end implicit denoising approach for respiratory sound classification, designed for datasets lacking dedicated denoising procedures. The method initially suppresses the majority of noise via the AFF technique, followed by a fine-grained noise reduction process leveraging differential attention in the DDL module. In addition, we introduced bias denoising loss and implemented combined training.  It further optimizes the model performance. Experimental results demonstrate the superior performance of our method across both CNN and Transformer architectures, exhibiting strong robustness and achieving a state-of-the-art accuracy of 65.53\%, surpassing all existing methods. In the future, we will continue to explore denoising techniques for the task of breath sound event detection.

\section{Acknowledgements}
This research is supported by the National Natural Science
Foundation of China under Grant No. 82260024. Acknowledgments to the First Affiliated Hospital of Nanchang University and the Laboratory of Imaging and Visual Representation of Nanchang University.

\bibliographystyle{IEEEtran}
\bibliography{mybib}

\end{document}